# A NOVEL LINEAR BATTERY ENERGY STORAGE SYSTEM (BESS) LIFE LOSS CALCULATION MODEL FOR BESS-INTEGRATED WIND FARM IN SCHEDULED POWER TRACKING

*Qiang Gui[1], Hao Su[1], Donghan Feng[1], Yun Zhou[1*], Ran Xu[1], Zheng Yan[1], Ting Lei[2]*

[1] Key Laboratory of Control of Power Transmission and Conversion, Ministry of Education, Department of Electrical Engineering, Shanghai Jiao Tong University, 800 Dongchuan Rd., Shanghai, China
[2] Marketing Department, State Grid Shanghai Municipal Electric Power Company, 1122 Yuansheng Rd., Shanghai, China
*yun.zhou@sjtu.edu.cn

## Abstract

Recently, rapid development of battery technology makes it feasible to integrate renewable generations with battery energy storage system (BESS). The consideration of BESS life loss for different BESS application scenarios is economic imperative. In this paper, a novel linear BESS life loss calculation model for BESS-integrated wind farm in scheduled power tracking is proposed. Firstly, based on the life cycle times-depth of discharge (DOD) relation-curve, the BESS life loss coefficient for unit throughput energy with different state of charge (SOC) can be determined from the life cycle times-DOD relation-curve fitting function directly. Secondly, as unidirectional variation of SOC in a single time step, the BESS life loss can be calculated through integration of the life loss coefficient-SOC relation function. A linear BESS life loss calculation model is established through self-optimal piecewise linearization of the primitive function of the life loss coefficient-SOC relation function. Thirdly, the proposed life loss calculation model is incorporated in the BESS-integrated wind farm scheduled power tracking optimization. Case studies demonstrate that with the proposed method, the BESS life loss item can be incorporated in the optimization model effectively, and the scheduled power tracking cost of the BESS-integrated wind farm can be determined and optimized more comprehensively.

## 1 Introduction

Due to rapid increase in global energy consumption and the diminishing of fossil fuels, the customer demand for new generation capacities and efficient energy productions keeps rising [1]. Compared with traditional power resources, the wind power is a cost-effective, emission-free generation resource with great benefits. However, the uncertainty and variability of renewable generation created challenges for wind power integration into grid [2]. Recently, rapid development of battery technology makes it feasible to integrate renewable generations with battery energy storage system (BESS) for alleviating uncertainty and variability [3].

The integrations of wind farms and BESS (i.e. BESS-integrated wind farm) are widely implemented to provide auxiliary services for wind farms, e.g. frequency control [4], wind power fluctuation smoothing [5], and wind farm scheduled power tracking [6]. For different BESS application scenarios in the BESS-integrated wind farms, typical conditions for the BESS (e.g. charge/discharge efficiency and upper/lower state of charge (SOC) limit) are considered as constraints in the optimization models [5], and the life cycle loss is considered as the mainly operation cost of the BESS [4], [6]. The SOC of a BESS is its available capacity expressed as a percentage of its rated capacity [3]. As a key indicator for the BESS to evaluate its residual useful life, the life cycle times is defined as how many times BESS could discharge from the full SOC to the specific SOC and then charge from the specific SOC to the full SOC until it is out of use. For a newly installed BESS, the total life cycle times of the BESS are related to the depth of discharge (DOD). Through discharge-charge cycling experiments, the relation-curve between the DOD and the total life cycle times of BESSs with different battery types can be referred to [7]. And based on the BESS life cycle times-DOD relation-curve and the SOC varying curve during the BESS utilizing time horizon, the rain-flow algorithm is widely used to calculate the life cycle loss of the BESS during the utilizing time horizon [8]-[10].

In the general rain-flow algorithm, the SOC varying curve during the BESS utilizing time horizon is the given input, the rain-flow calculation method is utilized to decompose the SOC varying curve into equivalent half and full discharge-charge cycles with different DODs. With the BESS life cycle times-DOD relation-curve, the life cycle loss of the BESS during the study time horizon can be calculated by accumulating the life cycle loss of each equivalent discharge-charge cycle determined by the rain-flow algorithm [8]. In [11], an improved rain-flow algorithm is proposed to count the equivalent cycles with different DODs more precisely, and the computing time for each rain-flow calculation process is reduced significantly. The rain-flow algorithm is a matured method to calculate the BESS life cycle loss.

However, the known SOC varying curve of the BESS is the essential input for the rain-flow calculation method. In the optimization models of different BESS application scenarios in the BESS-integrated wind farms [4]-[6], the SOCs of the BESS at different control moments in the study time horizon are variables to be optimized. It is hard to incorporate the BESS life cycle loss item determined by the rain-flow algorithm in the optimization model. Simplification is utilized to consider the BESS life cycle loss in the optimization model. For the simplification methods in [11], [12], although the BESS life cycle loss is considered, the BESS life cycle loss item is not incorporated into the optimization model directly.

Based on the life cycle times-DOD relation-curve, the BESS total throughput energy in discharge-charge cycles with different DOD can be derived from product of life cycle times and DOD in the relation-curve [13]. The BESS life loss can be determined by the ratio of throughput energy in the study time horizon to the total throughput energy [14]. In [14], the total throughput energy of the BESS is assumed as fixed value. And considering the impact of different SOC to the actual life loss, empirical BESS life loss coefficients with different SOCs are introduced to amend the throughput energy in the study time horizon. The BESS life loss is calculated through integration of the life loss coefficient-SOC relation function. To incorporate the BESS life loss item into the optimization model, a piecewise linearization method is introduced to process the primitive function of the life loss coefficient-SOC relation function.

The BESS life loss calculation model proposed in [14] can incorporate the BESS life loss item into the optimization model. However, for the proposed method in [14], as different BESS total throughput energy with different DOD can be derived from the life cycle times-DOD relation-curve directly [13], the fixed throughput energy assumption and the amendment of the throughput energy by the BESS life loss coefficients with different SOC are unnecessary, while the unmatched amendment will lead to big errors. And the non-optional empirical BESS life loss coefficients introduced in [14] (i.e. constant life loss coefficient with $SOC \in [0, 0.5]$ and linearly decreasing coefficient with $SOC \in (0.5, 1]$) are not suitable for BESSs with different battery technologies. Meanwhile, the life loss coefficient-SOC relation functions for BESSs with different battery technologies are not always monotonic decreasing, and the convexity of the primitive function of the life loss coefficient-SOC relation function may not be obtained. Therefore, the validity of the piecewise linearization method utilized in [14] will be affected.

In this paper, a novel linear BESS life loss calculation model for BESS-integrated wind farm in scheduled power tracking is proposed. The BESS life loss coefficient for unit throughput energy with different SOC is determined from the life cycle times-DOD relation-curve fitting function directly. Considering unidirectional variation of SOC in a single time step, the BESS life loss can be calculated through integration of the life loss coefficient-SOC relation function. The integration of the life loss coefficient-SOC relation function is transformed to the difference of the primitive function values. To incorporate the BESS life loss item into the optimization model, a revised self-optimal piecewise linearization (PWL) method is introduced to linearize the primitive function of the life loss coefficient-SOC relation function. The proposed BESS life loss calculation model is tested in the BESS-integrated wind farm scheduled power tracking optimization.

The major contributions of this paper are: 1) A novel mixed-integer linear BESS life loss calculation model is established. With the proposed method, the BESS life loss item can be incorporated into the optimization model of different BESS application scenarios effectively. 2) By incorporating the BESS life loss item, the scheduled power tracking cost of the BESS-integrated wind farm can be determined and optimized more comprehensively. The rest of the paper is organized as follows. The detailed BESS life loss calculation model is established in Section 2. With receding horizon optimization method, the BESS-integrated wind farm scheduled power tracking optimization model containing the BESS life loss item is presented in Section 3. Case studies to validate the proposed strategy are included in Section 4. Eventually, Section 5 concludes the paper.

## 2 Mixed-integer linear BESS life loss calculation model

*2.1 BESS life loss calculation model based on the life cycle times-DOD relation-curve*

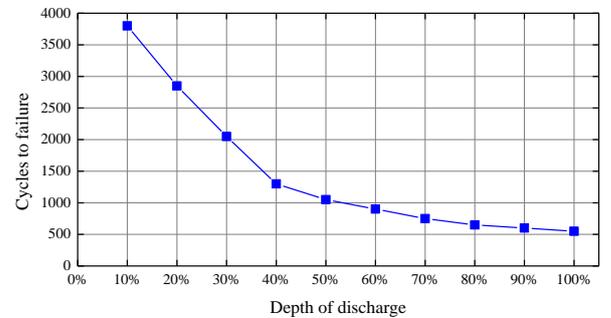

Fig. 1 Cycles to failure for a BESS with different DOD from experiment data supplied by the manufacturer

$$N_{\text{life}}(D_{\text{OD}}) = a_4 D_{\text{OD}}^4 + a_3 D_{\text{OD}}^3 + a_2 D_{\text{OD}}^2 + a_1 D_{\text{OD}} + a_0 \quad (1)$$

Fig.1 shows the cycles to failure for a typical BESS with different DOD from experiment data supplied by the manufacturer [13]. For BESSs with different battery technologies, the different life cycle times-DOD relation-curves can be fitted by different forms of functions (e.g. exponential functions or power functions) effectively. For experiment data in [13], as shown in (1), a four order function can be used to describe the life cycle times-DOD relation-curve [8], where $D_{\text{OD}}$ represents the DOD variable, $N_{\text{life}}(D_{\text{OD}})$ is the life cycle times with $D_{\text{OD}}$, $a_0 \sim a_4$ are fitting coefficients.

$$D_{\text{OD}} = 1 - S_{\text{OC}} \qquad (2)$$

$$L_{\text{loss}}(S_{\text{OC}}, 1) = \frac{1}{2N_{\text{life}}(1-S_{\text{OC}})} \qquad (3)$$

$$L_{\text{loss}}(S_{\text{OC}}, S_{\text{OC}}+\Delta S_{\text{OC}}) = L_{\text{loss}}(S_{\text{OC}}+\Delta S_{\text{OC}}, 1) - L_{\text{loss}}(S_{\text{OC}}, 1) \qquad (4)$$

$$\begin{aligned}\lambda_{\text{loss}}(S_{\text{OC}}) &= \lim_{\Delta S_{\text{OC}} \to 0} \frac{L_{\text{loss}}(S_{\text{OC}}, S_{\text{OC}}+\Delta S_{\text{OC}})}{\Delta S_{\text{OC}} C_{\text{rated}}} \\ &= -\frac{1}{2C_{\text{rated}}} \left( \frac{1}{N_{\text{life}}(1-S_{\text{OC}})} \right)' \end{aligned} \qquad (5)$$

The relationship between $D_{\text{OD}}$ and $S_{\text{OC}}$ is shown in (2). Based on the fitting function $N_{\text{life}}(D_{\text{OD}})$, the life loss for the BESS of the SOC variation interval $[S_{\text{OC}}, 1]$ can be calculated by (3). Considering a short SOC variation step $\Delta S_{\text{OC}}$, the short SOC variation interval $[S_{\text{OC}}, S_{\text{OC}}+\Delta S_{\text{OC}}]$ can be converted to the difference between interval $[S_{\text{OC}}, 1]$ and $[S_{\text{OC}}+\Delta S_{\text{OC}}, 1]$ equivalently, and the life loss for the BESS of the short SOC variation interval can be calculated by (4). On this basis, the BESS life loss coefficient for unit throughput energy with different SOC $\lambda_{\text{loss}}(S_{\text{OC}})$ can be determined from (5), where $C_{\text{rated}}$ is the rated BESS energy capacity. From the deriving results in (5), the life loss coefficient $\lambda_{\text{loss}}(S_{\text{OC}})$ can be determined by the fitting function $N_{\text{life}}(D_{\text{OD}})$ directly.

$$\begin{aligned} L_{\text{loss}}(t) &= \left| \int_{S_{\text{OC}}(t-1)}^{S_{\text{OC}}(t)} \lambda_{\text{loss}}(S_{\text{OC}}) d(S_{\text{OC}} C_{\text{rated}}) \right| \\ &= \left| \int_{S_{\text{OC}}(t-1)}^{S_{\text{OC}}(t)} \lambda_{\text{loss}}(S_{\text{OC}}) C_{\text{rated}} dS_{\text{OC}} \right| \end{aligned} \qquad (6)$$

For general BESS application scenarios, the SOC variation of the BESS in a single discrete time step is unidirectional. As shown in (6), the BESS life loss in the $t$th discrete time step $L_{\text{loss}}(t)$ can be calculated through integration of the life loss coefficient-SOC relation function, where $S_{\text{OC}}(t) / S_{\text{OC}}(t-1)$ is the SOC of the current/previous time step.

*2.2 Mixed-integer linear formulations of the BESS life loss calculation model*

$$L_{\text{loss}}(t) = \left| F(S_{\text{OC}}(t)) - F(S_{\text{OC}}(t-1)) \right| \qquad (7)$$

$$\begin{aligned} F(S_{\text{OC}}) &= \int_0^{S_{\text{OC}}} \lambda_{\text{loss}}(S_{\text{OC}}) C_{\text{rated}} dS_{\text{OC}} \\ &= \frac{1}{2} \left( \frac{1}{N_{\text{life}}(1)} - \frac{1}{N_{\text{life}}(1-S_{\text{oc}})} \right) \end{aligned} \qquad (8)$$

The BESS life loss calculation model in Section 2.1 is nonlinear, and the life loss item $L_{\text{loss}}$ in (6) is hard to be incorporated into the optimization model directly. By introducing the primitive function, the integration of the life loss coefficient-SOC relation function in (6) can be transformed to the difference of the primitive function values in (7), where the primitive function of the life loss coefficient-SOC relation function $F(S_{\text{OC}})$ is defined in (8).

$$f(y, \underline{y}, \overline{y}, \Lambda) = F(\underline{y}) + \sum_{\lambda=1}^{\Lambda} \phi_{y,\lambda} \Delta_{y,\lambda} \qquad (9)$$

$$\overline{\Delta}_y = (\overline{y} - \underline{y})/\Lambda \qquad (10)$$

$$\phi_{y,\lambda} = \frac{F(\underline{y}+\lambda\overline{\Delta}_y) - F(\underline{y}+(\lambda-1)\overline{\Delta}_y)}{\overline{\Delta}_y} \qquad (11)$$

$$\sum_{\lambda=1}^{\Lambda} \Delta_{y,\lambda} = y - \underline{y} \qquad (12)$$

$$\Delta_{y,\lambda} - \overline{\Delta}_y + (1-x_{\lambda+1})M + \varepsilon^+ \geq 0 \quad \forall \lambda=1,2,\ldots,\Lambda-1 \qquad (13)$$

$$0 \leq \Delta_{y,\lambda} \leq x_\lambda \overline{\Delta}_y \quad \forall \lambda=1,2,\ldots,\Lambda \qquad (14)$$

For BESSs with different battery technologies, the convexity of $F(S_{\text{OC}}(t)) - F(S_{\text{OC}}(t-1))$ cannot be obtained, a revised self-optimal PWL method proposed in [15] (shown in (9)-(14)) is introduced to linearize $F(S_{\text{OC}})$. In the left side of (9), $f(y, \underline{y}, \overline{y}, \Lambda)$ is the mixed-integer PWL function to linearize $F(S_{\text{OC}})$, where $y$ is the independent variable, $\underline{y} / \overline{y}$ is the lower/upper bound of $y$, and $\Lambda$ is the number of discretization in the PWL function. In the right side of (9), $\Delta_{y,\lambda}$ is the value of the $\lambda$th segment in the PWL function, and $\phi_{y,\lambda}$ is the slope parameter of the $\lambda$th segment. The upper bound of $\Delta_{y,\lambda}$ is defined in (10), and the detailed definition of $\phi_{y,\lambda}$ is presented in (11). In (12), the cumulative sum of the segments $\Delta_{y,\lambda}$ equals to the difference between $y$ and $\underline{y}$. By introducing a sufficiently big positive constant $M$, a near-zero positive constant $\varepsilon^+$, and a set of binary variables $\{x_\lambda\}$, the approximation error self-optimal conditions of the PWL function are shown in (13) and (14) [15].

$$L_{\text{loss}}(t) = \begin{cases} f(S_{\text{OC}}(t+1), S_{\text{OC}}^{\min}, S_{\text{OC}}^{\max}, \Lambda) - \\ f(S_{\text{OC}}(t), S_{\text{OC}}^{\min}, S_{\text{OC}}^{\max}, \Lambda) & v_{\text{ch}}(t)=1, v_{\text{dis}}(t)=0 \\ f(S_{\text{OC}}(t), S_{\text{OC}}^{\min}, S_{\text{OC}}^{\max}, \Lambda) - \\ f(S_{\text{OC}}(t+1), S_{\text{OC}}^{\min}, S_{\text{OC}}^{\max}, \Lambda) & v_{\text{ch}}(t)=0, v_{\text{dis}}(t)=1 \\ 0 & v_{\text{ch}}(t)=0, v_{\text{dis}}(t)=0 \end{cases} \qquad (15)$$

Replace $F(S_{\text{OC}})$ in (7) with $f(S_{\text{OC}}, S_{\text{OC}}^{\min}, S_{\text{OC}}^{\max}, \Lambda)$ and relax the absolute term by introducing the binary BESS charge and discharge state variables $v_{\text{ch}}(t)$ and $v_{\text{dis}}(t)$, the nonlinear BESS life loss calculation model in (7) is linearized to (15). As the condition constraint in (15) can be transformed into regular linear constraints conveniently, the BESS life loss item calculated by the mixed-integer linear BESS life loss calculation model in (15) can be incorporated into the optimization model of different BESS application scenarios effectively.

# 3 BESS-integrated wind farm scheduled power tracking optimization model

## 3.1 Scheduled power tracking optimization model containing BESS life loss item

As a typical BESS application scenario in the BESS-integrated wind farm, the BESS is integrated into wind farm to enable the BESS-integrated wind farm to inject energy into power grids as certain generation schedule determined previously, where the BESS is controlled to compensate stochastic power deviations between wind power and the predetermined generation schedules [6]. Referring to the practical wind power and RTO/ISO market rules [16], in the wind farm scheduled power tracking, balance fees or penalties are requested for the energy deviations between real-time output and day-ahead scheduled output (e.g. PJM, NYISO, ISO-NE, etc.). And tolerance deviations are set in the PJM market, for self-scheduled wind farms, differentials less than the tolerance (e.g. 5% or 5 MW) incur no charges.

Considering the application of the BESS in the wind farm, for the BESS-integrated wind farm scheduled power tracking, under the condition of time-of-use (TOU) electricity price, the energy deviations between real-time output and scheduled output can be selectively compensated by the local BESS or charged by the balance fees or penalties according to the market rules. Incorporating the BESS life loss item calculated by the linear model proposed in Section 2, the optimization model of the BESS-integrated wind farm scheduled power tracking is formulated as follows.

A. Objective function

$$\min \sum_{t \in S_H} BESS_{\text{loss}}(t) + Penalty(t) \quad (16)$$

$$S_H(t_i) = \{t_i, t_i+1, ..., t_i + H/\Delta t - 1\} \quad (17)$$

$$BESS_{\text{loss}}(t) = C_{\text{BESS}} L_{\text{loss}}(t) \quad t \in S_H \quad (18)$$

$$Penalty(t) = \gamma_{\text{lower}} C_E(t) P_{\text{out-of-limit}}^{\text{lower}}(t) \Delta t + \gamma_{\text{upper}} C_E(t) P_{\text{out-of-limit}}^{\text{upper}}(t) \Delta t \quad t \in S_H \quad (19)$$

In two-stage market clearing model, the wind farm day-ahead scheduled output is cleared in the day-ahead market [17]. To minimize the balance fees or penalties for the BESS-integrated wind farm scheduled power tracking, the objective function of the optimization model is shown in (16), where different items in the objective function are defined in (17)-(19) in detail. The set of discrete time $S_H$ in the advance control horizon $H$ started from the time step $t_i$ is defined in (17), where $\Delta t$ is the time step width. The cost for the BESS in compensating the deviations $BESS_{\text{loss}}(t)$ can be determined by (18), where $C_{\text{BESS}}$ is the total investment and operation cost of BESS, $L_{\text{loss}}(t)$ is the BESS life loss calculated by the mixed-integer linear BESS life loss calculation model in (15). Referring to the market rules [16], [18], the penalties for the wind farm in scheduled power tracking can be referred to (19), where $P_{\text{out-of-limit}}^{\text{lower}}(t)$ / $P_{\text{out-of-limit}}^{\text{upper}}(t)$ is the out of lower/upper tolerance limit values, $C_E(t)$ is the TOU electricity price, and $\gamma_{\text{lower}} / \gamma_{\text{upper}}$ is the out of lower/upper tolerance limit penalty coefficient.

B. Constraints

$$P_{\text{joint}}(t) = P_{\text{wind}}^f(t) + P_{\text{BESS}}(t) \quad t \in S_H \quad (20)$$

$$P_{\text{BESS}}(t) = P_{\text{BESS}}^{\text{dis}}(t) - P_{\text{BESS}}^{\text{ch}}(t) \quad t \in S_H \quad (21)$$

$$0 \leq P_{\text{BESS}}^{\text{dis}}(t) \leq P_{\text{BESS}}^{\text{dis,max}} v_{\text{dis}}(t) \quad t \in S_H \quad (22)$$

$$0 \leq P_{\text{BESS}}^{\text{ch}}(t) \leq P_{\text{BESS}}^{\text{ch,max}} v_{\text{ch}}(t) \quad t \in S_H \quad (23)$$

$$v_{\text{dis}}(t) + v_{\text{ch}}(t) \leq 1 \quad t \in S_H \quad (24)$$

$$S_{\text{OC}}(t) = S_{\text{OC}}(t-1) - \frac{\eta_{\text{dis}} P_{\text{BESS}}^{\text{dis}} \Delta t}{C_{\text{rated}}} + \frac{\eta_{\text{ch}} P_{\text{BESS}}^{\text{ch}} \Delta t}{C_{\text{rated}}} \quad t \in S_H \quad (25)$$

$$S_{\text{OC}}^{\min} \leq S_{\text{OC}}(t) \leq S_{\text{OC}}^{\max} \quad t \in S_H \quad (26)$$

$$P_{\text{out-of-limit}}^{\text{lower}}(t) = \max\{(1-\lambda_{\text{lower}})P_{\text{sch}}(t) - P_{\text{joint}}(t), 0\} \quad t \in S_H \quad (27)$$

$$P_{\text{out-of-limit}}^{\text{upper}}(t) = \max\{P_{\text{joint}}(t) - (1+\lambda_{\text{upper}})P_{\text{sch}}(t), 0\} \quad t \in S_H \quad (28)$$

The joint output of the BESS-integrated wind farm $P_{\text{joint}}(t)$ is constrained with the short-term wind farm forecast output $P_{\text{wind}}^f(t)$ and the BESS output $P_{\text{BESS}}(t)$ in (20). The BESS related constraints are shown in (21)-(26). In (21), $P_{\text{BESS}}^{\text{dis}}(t) / P_{\text{BESS}}^{\text{ch}}(t)$ is the discharge/charge power of the BESS. In (22) and (23), $P_{\text{BESS}}^{\text{dis,max}} / P_{\text{BESS}}^{\text{ch,max}}$ is the BESS discharge/charge power upper bound, and $v_{\text{dis}}(t) / v_{\text{ch}}(t)$ is the binary discharge/charge state variable. Constraint (24) will avoid the BESS to be operated in charging and discharging modes simultaneously. The SOCs of the BESS between two adjacent discrete time steps $S_{\text{OC}}(t-1)$ and $S_{\text{OC}}(t)$ are constrained in (25), where $\eta_{\text{dis}} / \eta_{\text{ch}}$ is the discharge/charge efficiency parameter. In (26), $S_{\text{OC}}^{\min} / S_{\text{OC}}^{\max}$ is the lower/upper bound of $S_{\text{OC}}(t)$. The out of scheduled power tracking lower/upper tolerance limit values $P_{\text{out-of-limit}}^{\text{lower}}(t)$ and $P_{\text{out-of-limit}}^{\text{upper}}(t)$ are determined by (27) and (28) respectively, where $P_{\text{sch}}(t)$ is the wind farm day-ahead scheduled output, and $\lambda_{\text{lower}} / \lambda_{\text{upper}}$ is the lower/upper tolerance deviation rate parameter for the joint output $P_{\text{joint}}(t)$ derived from the scheduled output $P_{\text{sch}}(t)$.

$$P_{\text{joint}}(t) + P_{\text{out-of-limit}}^{\text{lower}}(t) \geq (1-\lambda_{\text{lower}})P_{\text{sch}}(t) \quad (29)$$

$$P_{\text{joint}}(t) - P_{\text{out-of-limit}}^{\text{upper}}(t) \leq (1+\lambda_{\text{upper}})P_{\text{sch}}(t) \quad (30)$$

$$P_{\text{out-of-limit}}^{\text{lower}}(t) \geq 0 \quad (31)$$

$$P_{\text{out-of-limit}}^{\text{upper}}(t) \geq 0 \quad (32)$$

For constraints in (20)-(28), only (27) and (28) are nonlinear constrains. Considering the minimization of the penalty item $Penalty(t)$ in the objective function in (16), constraints (27) and (28) can be substituted by the linear constraints (29)-(32) equivalently. Considering the BESS life loss item incorporated into the objective function, the BESS life loss calculation model in (15) is also included as constraint in the optimization. Therefore, the BESS-integrated wind farm scheduled power tracking optimization model is with objective function defined in (16)-(19), and constraints in (15), (20)-(26), and (29)-(32). As all constraints are linear constraints with binary variables, the formulated BESS-integrated wind farm scheduled power tracking optimization model is a mixed-integer linear programming (MILP) model, and can be solved by commercial solvers effectively.

## 3.2 Receding horizon optimization process for the scheduled power tracking optimization model

For the scheduled power tracking optimization model established in Section 3.1, the short-term wind farm forecast output $P_{wind}^f(t)$ in the advance control horizon $H$ are utilized in the optimization model. Take the latest national standard of China for example, for short-term wind power forecast (timescale less than 4h), maximal short-term forecast error limitation is requested, and the accuracy of the short-term wind power forecast output can be obtained [3]. In accordance with the usage of short-term wind power forecast output to maintain the accuracy of the optimized results, a shorter advance control horizon $H$ (shorter than the short-term forecast timescale) is considered in the optimization model. And the matured receding horizon optimization process [19] is used to determine the real-time scheduled power tracking optimization results for the BESS-integrated wind farm to meet the real-time application requirements.

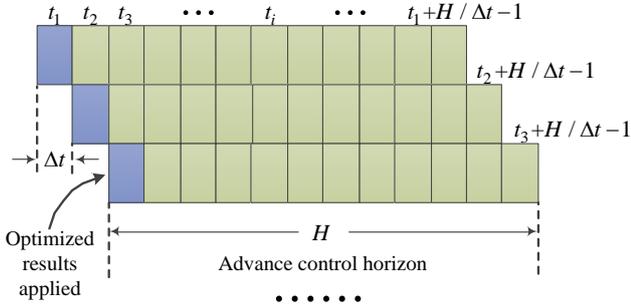

Fig. 2 Time-line of the receding horizon optimization process of the scheduled power tracking optimization model

Referring to the receding horizon optimization steps in [19], the time-line of the receding horizon optimization process of the scheduled power tracking optimization model is shown in Fig. 2, and the detailed steps for the receding horizon optimization process of the scheduled power tracking optimization model are as follows:

1) Step 1: At a given $t_i$th discrete time step, collect parameters for the scheduled power tracking optimization model, including $C_E(t)$, $P_{wind}^f(t)$, $P_{sch}(t)$ in $H$, and $S_{OC}(t_i-1)$ in the $t_i-1$th time step etc. Solve the corresponding optimization model established in Section 3.1.

2) Step 2: In the solving results of the control variables in the optimization model in $H$ (i.e. $v_{dis}(t)$, $v_{ch}(t)$, $P_{BESS}^{dis}(t)$, $P_{BESS}^{ch}(t)$, $P_{out\text{-}of\text{-}limit}^{lower}(t)$, $P_{out\text{-}of\text{-}limit}^{upper}(t)$, and $P_{joint}(t)$, etc. $t \in S_H(t_i)$), the solving results of the control variables of the first time step in $H$ (i.e. $v_{dis}(t_i)$, $v_{ch}(t_i)$, $P_{BESS}^{dis}(t_i)$, $P_{BESS}^{ch}(t_i)$, and $P_{joint}(t_i)$ etc.) are applied as the real-time scheduled power tracking control results of the current $t_i$th discrete time step.

3) Step 3: For the following $t_i+1$th discrete time step, repeat Step 1 and 2.

It's important to note that for the scheduled power tracking starting discrete time step $t_1$, an initial SOC of the BESS $S_{OC}(0)$ is needed and set as given value. Through receding horizon optimization process in Step 1-3, the scheduled power tracking results for the BESS-integrated wind farm can be optimized and the real-time application requirements for the BESS-integrated wind farm scheduled power tracking can be satisfied meanwhile.

## 4 Case study

The proposed BESS-integrated wind farm scheduled power tracking optimization model with receding horizon optimization process is tested on an offshore wind farm (165 MW installed capacity) in Estinnes, Belgium. A lithium iron phosphate (LFP) BESS is allocated in the wind farm to track the day-ahead wind farm scheduled output. Fig. 3 illustrates one typical day's actual wind power day-ahead forecast output of the test wind farm [20]. As a typical method, the day-ahead wind farm scheduled output $P_{sch}(t)$ utilized in the case study is generated by the day-ahead forecast output with an hourly moving-average filter (shown in Fig. 3).

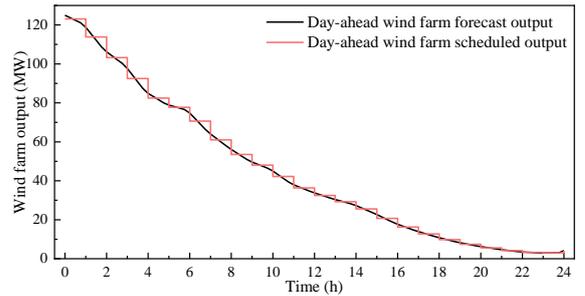

Fig. 3 Day-ahead forecast output and scheduled output on a typical day of the test wind farm in Estinnes, Belgium

Except for the day-ahead wind farm scheduled output, as necessary input of the proposed optimization model, the actual intra-day most recent forecast output of the wind farm at the same typical day [20] is utilized as the short-term wind farm forecast output $P_{wind}^f(t)$, while a typical hourly data for TOU electricity price of Belgium $C_E(t)$ is taken from [21]. And the curves of $P_{wind}^f(t)$ and $C_E(t)$ used in the case study are shown in Fig. 4.

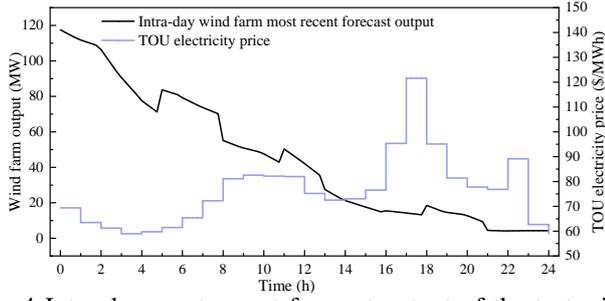

Fig. 4 Intra-day most recent forecast output of the test wind farm and corresponding TOU electricity price

$$N_{\text{life}}(D_{\text{OD}}) = 49660e^{-14.32D_{\text{OD}}} + 34280e^{-2.181D_{\text{OD}}} \quad (33)$$

Table 1 Detailed parameters of the LFP BESS in the BESS-integrated wind farm

| Parameter | Value | Parameter | Value |
|---|---|---|---|
| $C_{\text{rated}}$ | 25 MWh | $C_{\text{BESS}}$ | $\$1.285 \times 10^7$ |
| $P_{\text{BESS}}^{\text{dis,max}}$ | 10 MW | $P_{\text{BESS}}^{\text{ch,max}}$ | 10 MW |
| $\eta_{\text{ch}}/\eta_{\text{dis}}$ | 0.95/1.05 | $S_{\text{OC}}^{\min}/S_{\text{OC}}^{\max}$ | 0.15/0.85 |

For the LFP BESS considered in the BESS-integrated wind farm, the different life cycle times-DOD relation-curve experiment data of the LFP BESS is referred to [7]. Shown in (33), the sum of two exponential functions is introduced in [7] to fit the life cycle times-DOD relation-curve of the LFP BESS. And the detailed parameters of the LFP BESS are tabulated in Table 1.

Table 2 Other parameters of the scheduled power tracking optimization model with receding horizon optimization process

| Parameter | Value | Parameter | Value |
|---|---|---|---|
| $\Lambda$ | 10 | $\lambda_{\text{lower}}/\lambda_{\text{upper}}$ | 0.05 |
| $\gamma_{\text{lower}}/\gamma_{\text{upper}}$ | 1.0 | $H$ | 2 h |
| $\Delta t$ | 15 min | $S_{\text{OC}}(0)$ | 0.5 |

The other essential parameters of the BESS-integrated wind farm scheduled power tracking optimization model with receding horizon optimization process are shown in Table 2. For the BESS life loss calculation model established in Section 2.2, the number of discretization $\Lambda$ is set as 10 in (15). The lower/upper tolerance deviation rate parameter $\lambda_{\text{lower}}/\lambda_{\text{upper}}$ and out of lower/upper tolerance limit penalty coefficient $\gamma_{\text{lower}}/\gamma_{\text{upper}}$ are set for 0.05 and 1.0 respectively. In the receding horizon optimization process, $H$ is set for 2 h, and in accordance with the time step in the input data, $\Delta t$ is set for 15 min. The initial SOC of the BESS $S_{\text{OC}}(0)$ is set for 0.5. It's important to note that considering the advance control horizon $H$ in the receding horizon optimization process, extra 2 hours data of $P_{\text{sch}}(t)$, $P_{\text{wind}}^{\text{f}}(t)$, and $C_{\text{E}}(t)$ shown in Fig. 3 and Fig. 4 are needed in the case study.

The proposed BESS-integrated wind farm scheduled power tracking optimization model with receding horizon optimization process is programmed and realized on the MATLAB platform with the YALMIP toolbox and the GUROBI is the MILP solver. With the input data and parameters in Fig. 3, Fig. 4, Eq. (33), Table 1, and Table 2 etc., the solving results of the BESS-integrated wind farm scheduled power tracking optimization model with receding horizon optimization process for the test BESS-integrated wind farm (i.e. the 165 MW offshore wind farm in Estinnes, Belgium with 25MWh LFP BESS) are shown in Fig. 5 and Table 3.

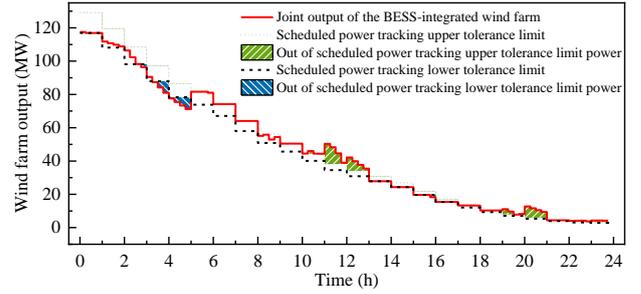

(a) Scheduled power tracking results of the BESS-integrated wind farm

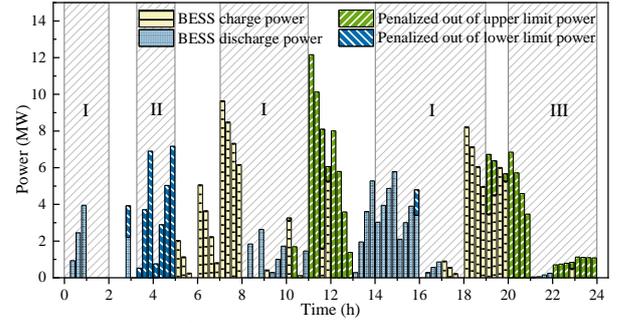

(b) BESS discharge/charge power and out of scheduled power tracking lower/upper tolerance limit power

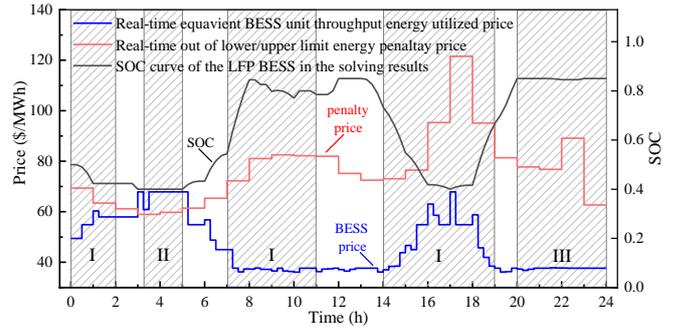

(c) BESS SOC curve and real-time price curve for the out of scheduled power tracking tolerance limit energy compensated by the BESS/penalized by the market rules

Fig. 5 Schematic diagrams of the solving results of the BESS-integrated wind farm scheduled power tracking

In Fig. 5(a), the scheduled power tracking lower/upper tolerance limit is generated by the day-ahead wind farm scheduled output $P_{\text{sch}}(t)$ with the lower/upper tolerance deviation rate parameter $\lambda_{\text{lower}}/\lambda_{\text{upper}}$ in Table 2. If the optimized joint output is within the lower and upper limits, only BESS life loss cost will be created when the BESS is utilized to compensate the deviations between wind power and the lower/upper tolerance limit. On the other hand, for lack of compensation by the BESS due to expensive BESS life loss price or none discharge/charge capacity of the BESS (i.e. BESS SOC reaches the lower/upper bound $S_{OC}^{min}$ / $S_{OC}^{max}$), when threshold-crossing happens, the out of lower/upper limit part will be penalized by the market rules shown in (19). The detailed discharge/charge power of the BESS and the out of lower/upper limit power are shown in Fig. 5(b). And the SOC curve of the BESS, the real-time equivalent BESS unit throughput energy utilized price (determined by the corresponding $S_{OC}(t)$ and the discharge/charge tendency of the BESS), and the real-time out of lower/upper limit energy penalty price are shown in Fig. 5(c).

The objective function value and other key index values of the solving results of the BESS-integrated wind farm scheduled power tracking are listed in Table 3. The total cost for the test BESS-integrated wind farm in scheduled power tracking of the typical day is $ 3861.45, including the BESS life loss cost $ 1769.55 and the penalty due to out of lower/upper limit energy charged by the market rules $ 2091.90. The total discharge/charge throughput energy of the BESS in scheduled power tracking and total out of lower/upper tolerance limit energy are 41.73 MWh and 28.37 MWh respectively.

Table 3 Objective function value and other key index values of the solving results of the BESS-integrated wind farm scheduled power tracking

| Index | Value |
|---|---|
| Total cost of scheduled power tracking (i.e. objective function) | $ 3861.45 |
| Total discharge/charge throughput energy of the BESS | 41.73 MWh |
| Total BESS life loss cost in scheduled power tracking | $ 1769.55 |
| Total out of lower/upper tolerance limit energy | 28.37 MWh |
| Total out of limit penalty charged by the market rules in scheduled power tracking | $ 2091.90 |

Further analysis regarding the behaviors of the BESS in scheduled power tracking from solving results in Fig. 5 is as follows. For demo time durations I in Fig. 5, shown in Fig. 5(c), the equivalent BESS unit throughput energy utilized price (BESS price in short) is cheaper than the real-time out of lower/upper limit energy penalty price (penalty price in short). Illustrated in Fig. 5(b), in demo time durations I, the BESS is discharged or charged frequently to compensate the deviations between wind power and the lower/upper tolerance limit, and near to zero out of lower/upper limit penalty is charged by the market rules.

For demo duration II, shown in Fig. 5(c), the BESS price is more expensive than the penalty price, the deviations between the real-time output and the lower/upper limit is selectively to be penalized by the market rules to minimize the objective function value (i.e. sum of the BESS life loss cost and the out of limit penalty by the market rules shown in (16)). For demo time duration III, although the BESS price is cheaper, the BESS SOC reaches the upper bound $S_{OC}^{max}$ shown in Fig. 5(c). And in demo time duration III, the real-time wind farm output is higher than the scheduled output tolerance upper limit shown in Fig. 5(a), as none charging capacity for the BESS when SOC reaches the upper bound, the deviations between the real-time output and the upper limit in demo time duration III are penalized by the market rules without choice. Therefore, the comparisons of the behaviors of the BESS in demo duration I, II and III demonstrate the BESS life loss item is incorporated in the BESS-integrated wind farm scheduled power tracking optimization model effectively.

$$\min \sum_{t \in S_H(t_1)} Penalty(t) \quad (34)$$

Table 4 Objective function values of the solving results of the BESS-integrated wind farm scheduled power tracking optimization models with different objective functions

| No. | Case 1 | Case 2 |
|---|---|---|
| Objective function | (16) | (34) |
| Objective function value | $ 3861.45 | $ 920.23 |
| Total out of limit penalty | $ 2091.90 | $ 920.23 |
| Total BESS life loss cost | $ 1769.55 | $ 5162.58 |
| Sum of BESS life loss cost and the out of limit penalty | $ 3861.45 | $ 6082.81 |

Table 4 summarizes the detailed function values of the solving results of the BESS-integrated wind farm scheduled power tracking optimization models with different objective functions. With the objective function in (16), the Case 1 is just the same case with solving results shown in Fig. 5 and Table 3. And the second column results of Case 1 in Table 4 are same to the results in Table 3.

Removing the BESS life loss item in the objection function in (16), the objective function in (34) is considered in Case 2 in Table 4. Comparing the results in Table 4, with a simplified objective function, the objective function value of Case 2 illustrates much smaller to Case 1. However, the actual total BESS life loss cost of the BESS-integrated wind farm scheduled power tracking of Case 2 is $ 5162.58, which leads to a much higher sum of BESS life loss cost and the out of limit penalty compared to Case 1. The comparisons of the results of Case 1 and 2 in Table 4 demonstrate that through incorporating the BESS life loss item in the optimization model, the scheduled power tracking cost of the BESS-integrated wind farm can be determined and optimized more comprehensively.

## 5 Conclusion

This paper presents a linear BESS life loss calculation model to determine the BESS life loss in the BESS utilizing time horizon. The BESS life loss is calculated through integration of the life loss coefficient-SOC relation function developed from the life cycle times-DOD relation-curve fitting function directly. Considering the integration of the life loss coefficient-SOC relation function is equivalent to the difference of the primitive function values, the mixed-integer linear BESS life loss calculation model is established through self-optimal PWL of the primitive function of the life loss coefficient-SOC relation function. With the proposed BESS life loss calculation model, the BESS life loss item can be incorporated into the optimization model of different BESS application scenarios.

The proposed BESS life loss calculation model is tested in the BESS-integrated wind farm scheduled power tracking optimization in this paper. Detailed scheduled power tracking optimization model containing BESS life loss item is established, and the receding horizon optimization process is introduced to solve the optimization model. Case studies on a 165 MW offshore wind farm in Estinnes, Belgium with 25MWh LFP BESS demonstrate that with the proposed method, the BESS life loss item can be incorporated in the optimization model effectively, and the scheduled power tracking cost of the BESS-integrated wind farm can be determined and optimized more comprehensively.

# 6  Acknowledgements

This work was sponsored and supported by National Key R&D Program of China under Grant 2017YFB0902200, Science and Technology Project of State Grid Corporation of China under Grant 5228001700CW, National Natural Science Foundation of China under Grant 51907122, Shanghai Sailing Program under Grant 19YF1423800, and Science and Technology Project of State Grid Shanghai Municipal Electric Power Company of China under Grant SGSH0000YXJS1900181.